\documentclass[aps,twocolumn]{revtex4}
\usepackage{graphicx}
\usepackage{dcolumn}
\newcommand{\be}{\begin{equation}} 
\newcommand{\ee}{\end{equation}} 
\newcommand{\bea}{\begin{eqnarray}} 
\newcommand{\eea}{\end{eqnarray}} 
\newcommand{\bc}{\begin{center}} 
\newcommand{\ec}{\end{center}}

\begin{document}
\bibliographystyle{plain}
\pagenumbering{arabic} 

\title{Evolutionary route to diploidy and sex}

\author{Erkan T{\"u}zel$^{1,2}$, Volkan Sevim$^{1}$ and Ay{\c s}e 
Erzan$^{1,3}$}

\affiliation{$^1$ Department of Physics, Faculty of Sciences and Letters\\
Istanbul Technical University, Maslak 80626, Istanbul, Turkey \\
$^2$ Department of Physics, Faculty of Sciences and Letters \\
        I{\c s}{\i }k University, Maslak, 80670, Istanbul, Turkey\\
$^3$ G\"ursey Institute, P.O.B. 6, \c Cengelk\"oy, 81220 Istanbul, Turkey
}

\begin{abstract} 
Using a bit-string  model of evolution,
we find a successful route to diploidy and sex in simple
organisms.  Allowing the sexually reproducing diploid individuals to 
also perform mitosis, as they do in a haploid-diploid cycle, leads to 
the complete takeover of the population by sexual diploids. This mechanism is 
so robust, that even the accidental conversion and  pairing  of
only two diploids give rise to a sexual population.

\end{abstract}

\maketitle

\section{Introduction} 

The evolution of sex through Darwinian selection in spite of the seeming odds 
with respect to asexual reproduction through cloning, is a still unresolved 
problem~\cite{Smith1,Smith2,Stearns}, which has attracted a great deal of recent 
activity~\cite{Science,Drossel}. We would like to present here results from a 
very simple model based on unicellular organisms where sex seems to have first 
emerged, without any complicating factors such as the cost of male-female 
differentiation, parental care, mating preferences, maternity periods and so on, 
which we assume only to have arisen in higher organisms. These factors may be 
responsible for the differentiation, modification or even at times reversal (as 
in meiotic parthenogenesis~\cite{Smith2},p.234) of the patterns that may have 
become accidentally established in the remote evolutionary past.  Unicellular 
organisms are not only simpler to model, but this choice also imposes more 
stringent conditions on the establishment of a sexual population, since it 
severly limits the number of viable offspring that a pair of parent cells can 
have. By confining ourselves to ``worst case scenarios" we hope to be able to 
obtain plausible lower bounds to the feasibility of sex as a reproductive 
mechanism.

Sexual reproduction 
as we know it today in higher plants and animals 
can be regarded as a 
haploid-diploid cycle with a highly abbreviated haploid phase, where the 
(haploid) gametes typically do not perform mitosis.~\cite{Smith1,Smith2} 
These haploids are not viable unless
two of them fuse and once more make a diploid cell, which then multiplies by 
mitosis and eventually forms the mature individual.

The haploid diploid cycle (HDC) where, under unfavorable conditions the 
population 
becomes diploid, 
 is found in  many species~\cite{Candan1,Candan2}, and was the motivation 
behind the 
hypothesis of Jan, Stauffer and Moseley~\cite{Jan1} (JSM) who proposed that 
diploidy and 
sex may have first arisen as a way 
to escape death, when a simple, unicellular  individual is threatened by too 
many deleterious 
mutations. The fitness is taken to be a step function of the number of 
mutations, also setting the threshold for conversion to diploidy and sex. Here, 
the number of 
deleterious mutations can be read as the distance from the ideal case, called 
the 
wildtype, which can be altered as a result of a change in the environment, 
so  that the individual is not as well 
adapted as before the change.  

In previous work~\cite{Orcal,Tuzel}  we  showed that the JSM hypothesis indeed 
leads to a steady source of diploid sexuals, given a population of haploids 
which multiply by mitosis. This 
leads to a steady state distribution of coexisting haploids and diploids in a 
constant population. The premises 
adopted in~\cite{Tuzel} were very restrictive.  Two sexual ``parent" cells were 
allowed to have only one offspring, after
which they died. In case one considers greater number of offspring, by allowing 
a greater number of viable gametes to be 
formed, as is indeed possible in many unicellular 
organisms~\cite{Candan1}, one 
finds that the diploid, sexual population completely takes over.

In the same paper\cite{Tuzel}, we have also considered the situation where
conversion to sex occurred with a constant probability $\sigma$ (which we
varied between 0 and 1), independently of the fitness of the individual.  
This strategy also gave rise to a sexual population making up a small but
macroscopic fraction of the total, so that the JSM threshold mechanism for
conversion to sex did not prove to be necessary, although it was more
successful. On the other hand the JSM hypothesis has the additional
attraction of providing a mechanism which could trigger the fusion of two
haploid cells to form a diploid:  it is known~\cite{Smith1},p.149, that 
extensive genetic
damage can lead to gene repair via genetic transmission between two
haploid cells; the fusion of two haploids could be seen as an extreme form
of such behaviour.  An alternative means of forming a diploid from a
haploid cell is via endomitosis~\cite{Smith2},p.230, as in the first step 
in
meiotic parthenogenesis. Endomitosis is the process whereby the genetic
material in a cell is duplicated without subsequent cell division as in
normal mitosis.  Again, it is not implausible that this process originated
as a result of grave genetic damage which precluded the successful
completion of mitosis.~\cite{Smith1}, p.149.

In the present paper we show that, if the diploid cells, once formed, are
also allowed to multiply by mitosis, as indeed they do in a
haploid-diploid cycle, the whole population is taken over by diploid
cells, who perform facultative sex if they are once more threatened by
extinction due to too many deleterious mutations. Moreover we show that
even an episodic conversion to sex, involving as few as only two
individuals who survive to mate, leads to a steady state made up solely of
sexual types.

In the next section we briefly describe our algorithms and report the 
simulation results.  In section 3, we provide a discussion and 
some pointers for future research.

\section{Bit-string model for the Conversion to Sex - Algorithms and 
simulation results}

The type of model we consider here is the same as in Ref.~\cite{Tuzel}.  
Below and in Section A, we recall the definitions.
In Section B and C we introduce
new rules to test the autonomous viability of sexual populations.

Each haploid one-celled organism consists, for our purposes, of a 15-bit
string of ``0"s and ``1"s, representing the genetic code in a 16-bit
computational word. ~\cite{Muller,Stauffer,Penna} We use the bit defining
the ``sign", to specify whether the individual is asexual (+) or sexual
(-).  A mutation consists of flipping a randomly chosen bit except the
sign bit, with a constant probability, $\Gamma$, for each individual per
generation.  Since the genetic difference between individuals of the same
species is typically less than $10\%$~\cite{Smith1},p.52, this rather short
string for the genetic code may be considered as a coarse grained model
for the complete genome of the individual, which we divide up into
different zones, retaining a ``0" where there are no mutations, and
flipping the bit to ``1" if there are one or more mutations in this zone.  
The wild type
is a string of all ``0"s.  Therefore, at each locus, a ``1" corresponds to
a deleterious mutation (which we will call ``mutation," for short, where
this is not liable to lead to any confusion.) Diploid organisms have two
bit strings, which are allowed to be different. The number of deleterious
mutations $m$ is simply the number of ``1"s for a haploid individual. For
a diploid, the number of ``expressed" deleterious mutations is the number
of loci at which both homologous alleles are set to ``1," i.e., we assume
that deleterious mutations are recessive.

The total population is fixed at $N=10^3$, and we have chosen
$\Gamma=1/N$.  This corresponds to a mutation rate per allele per
generation of $\sim 6 \times 10^{-5}$, which is comparable to the typical
mutation rates encountered in eukaryotes~\cite{Smith2},p.64.

The probability of survival (or fitness) as a function of $m$ 
 is given by a Fermi-like
distribution~\cite{Thoms},
$P(m)$,
\be P(m) = { 1  \over {\exp[\beta (m - \mu)] + 1}} \;\;\;.
\label{Fermi} \ee
For large $\beta$ (or ``low temperatures," in the language of statistical
mechanics), $P(m)$ behaves like a step function. Individuals with $m>\mu$
die, those with $m<\mu$ survive, and those with $m=\mu$ survive with a
probability of $1/2$. In the simulations we confined ourselves to low
temperatures ($\beta=10$). We chose $\mu=4$ which just allows us enough
variability without leading to totally unrealistic mutational loads.

We start with a set of $N$ asexual (haploid) wildtypes.
In each generation about $\Gamma N$ individuals suffer mutations; they are 
killed off or retained according to the fitness function~(\ref{Fermi}), and 
the population restored to $N$ by once cloning as many randomly chosen survivors 
as necessary.
The population
of haploid asexuals settles down to a minimally stable~\cite{Chisolm} steady 
state 
distribution~(see Table I) as shown in Fig. 1,
independently of the value of $\Gamma$, for $\Gamma \geq 
1/N$.~\cite{Orcal,Tuzel}

For comparison, we have also performed simulations on a diploid
population reproducing asexually, according to the same rules as stated in
the previous paragraph. We have found that the diploid population reaches
a steady state with an $m$ distribution peaked at $m=2$, rather than 
$m=3$ as found for haploids.
(See Fig. 1 and Table I).

\subsection{Conversion to sex}

Let us briefly summarise our algorithm for conversion to sex, which we
also used in our previous study (see Model A, Ref.~\cite{Tuzel}).

We choose the steady state of the haploid population as our initial state
from which to start the conversion to sex, as that would be a most likely
``natural state" encountered at this stage. Once the haploid population
reaches a steady state, we allow those individuals that are threatened by
extinction to convert to diploidy and sex, by implementing the following
rule at each generation: If an asexual individual with $m=4$ has survived,
it is converted to an active sexual, by deterministically and irreversibly
switching its sign bit to (-). It performs endomitosis and becomes
diploid. If there are already sexual, diploid organisms in the population,
they will also be made active if $m$, the number of their expressed
mutations, is $\ge 4$ (otherwise they do not participate in the
reproduction cycle; hence the conversion to sex may be termed
non-hereditary).  Finally, all the active sexual organisms pair randomly
and engage in sexual reproduction, where they each contribute one gamete
(formed via one step meiosis)  towards a single diploid sexual offspring.
If we denote the genotypes of parents as $\{A a\}$ and $\{B b\}$
respectively, then the genotype of the offspring is either $\{AB\}$,
$\{Ab\}$, $\{aB\}$ or $\{ab\}$. No crossover occurs during this one step
meiosis. If there is only one active sexual at a certain time step then it
must wait subsequent generations until it either finds a partner or it
dies.  We keep the population constant~\cite{Redfield} by cloning randomly
selected {\em haploid} individuals to make up the deficit.

(The consequences of the conversion to hereditary and obligatory sex can
be found in Ref.~\cite{Tuzel}.)

\subsection{\bf Mitotic Diploid  
Sexuals Win over Haploids}

In the previous subsection we described a scenario where 
too big a mutation load meant conversion to diploidy and sex, if the individual 
survived. It should be realised that in this scheme 
the haploids, who multiply by cloning, provide a steady source for the diploid 
sexuals, 
whose numbers are halved every time they mate. 
Now we want to test whether the diploid, sexual population can 
survive autonomously, if  the sexuals  are also allowed to 
perform mitosis themselves.

Once a steady state with coexisting haploid and diploid populations is
reached via the algorithm described above, we switch off the conversion of
haploid organisms to diploidy and sex.  The diploid individuals mate when
they face extinction due to too many mutations.
 We now keep the total population
constant by making up for the deficit in the population at each step by
cloning randomly selected individuals, {\em regardless of whether they are
sexual or asexual.} In colonies undergoing a haploid-diploid cycle, it is
 quite frequently the case~\cite{Candan1} that the diploid phase of the cycle 
also involves multiplication by mitosis.  
 
The result is that the diploid, sexual individuals completely win over the
population. The haploids which now can not compete with the sexuals, become
extinct. Here we find that the haploid phase of the HDC becomes
abbreviated to the point where haploids appear only as gametes which do
not perform mitosis.  This is exactly the situation in highly evolved
sexual organisms.

%%Figure 1

\begin{figure}[!ht]
\leavevmode
\rotatebox{270}{\scalebox{.4}{\includegraphics{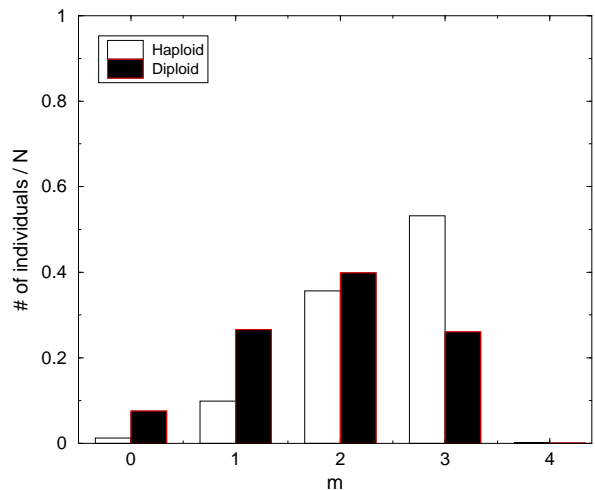}}}
\caption{
 The steady state distribution of pure haploids with respect to the number
deleterious mutations $m$, compared with the distribution of asexual diploids
over the number of {\em expressed} mutations $m$. See text. The histograms 
are normalized to unity. 
}
\end{figure}       

\subsection{Episodic conversion to diploidy and sex leads to sexual steady
state}

One can also think of a scenario in which conversion to sex takes place
accidentally over a short period of time after the asexual population
settles down to a steady state. In fact this is probably the most
realistic situation, given the random nature of the mutations.

\begin{table*}
\begin{ruledtabular}
\caption{The distribution of expressed deleterious mutations and death rates for 
the models discussed in Section II.
The $m$-distributions are percentages ($\pm 1 \%$) computed over each type 
if coexisting; the death rates are per capita per generation. ($^*$ 
see Ref.~\cite{Tuzel})}
\begin{tabular}{cc|c|c|c|c|c|c}
                         &          & $m=0$ & $m=1$ & $m=2$ & $m=3$ & $m=4$ & 
death rate \\ \colrule
Pure Haploid             &          & 1     & 10    & 36    & 53    & 0     & 
$O(10^{-4})$      \\
Pure Diploid             &          & 7     & 27    & 40    & 26    & 0     & 
$O(10^{-5})$      \\
Diploid Sexuals          & Mitotic  & 7     & 27    & 40    & 26    & 0     & 
$O(10^{-5})$      \\
Diploid Sexuals          & Episodic & 7     & 27    & 40    & 26    & 0     & 
$O(10^{-5})$      \\
Model A - non-hereditary$^*$  & Asex     & 1     & 11    & 37    & 51    & 
0     & $O(10^{-4})$      \\
Model A - non-hereditary$^*$  & Sex      & 11    & 33    & 36    & 20    & 
0     & $O(10^{-4})$      \\
Model A - hereditary$^*$     & Asex     & 1     & 11    & 36    & 52    & 
0     & $O(10^{-4})$      \\
Model A - hereditary$^*$     & Sex      & 10    & 50    & 40    & 0     & 
0     & $O(10^{-2})$      \\ 
\end{tabular}
\end{ruledtabular}
\end{table*}

The way we actually implemented this in the computer code was by 
deterministically switching the sign bits of the first two asexual individuals 
to survive with $m=4$, and then turning off the possibility of further 
conversion.
These then form two sexual individuals by  endomitosis, and if the first 
survives long 
enough so that it can mate with the second, will give rise to one 
sexual offspring. 

The rest of the rules are as explained before; in each generation we clone
randomly chosen individuals to make up the deficit in the population,
regardless of whether they are asexuals or sexuals.  We allow the diploid
individuals to mate when they face extinction in the course of their
lives.

Suprisingly, the sexuals capture the population in 95 percent of the
performed runs. (In the rest, the single diploid, which is still at the
threshold of extinction with an $m$ value of 4, may not survive until a
partner arrives.)

\subsection{Relative Fitness}

The steady state $m$ distributions for the sexual populations treated in Section 
II.B and C are much better than the haploid distribution. However, they are 
identical with the distribution for asexual diploids.(see Table I and Fig. 1)
This means that as soon as diploidy is achieved, it is so
successful in screening the effects of deleterious mutations that once
diploid, the organism practices sex very infrequently, making these three
distributions identical.

The diploid $m$-distribution in Fig. 1 should be contrasted with the
results of Ref.~\cite{Tuzel} (Model A, non-hereditary and hereditary; 
also see Subsection A, above) where cloning of the
sexuals is not allowed, but there is a steady influx of new sexuals from
the haploids (see Table I). In the non-hereditary model, the $m$
distribution has shifted nearer the wildtype than in the purely diploid
case (although the peak is again at $m=2$); in the hereditary model, the
peak has shifted to $m=1$.  Thus, the frequency with which sex is
practiced has a salutary effect on the distribution of expressed
deleterious mutations in the population.

On the other hand it should be noted that, with the Fermi-like fitness
function (Eq.(1)) for large $\beta$, to which we have confined ourselves,
the survival probability is $\sim$ constant for $m <\mu$, and therefore
does not discriminate between steady state populations that differ solely
in the shape of their $m$ distributions for $m <\mu$. A better measure
might be the deathrate, namely, the average number of individuals which are 
eliminated at each step, under the criterion given in Eq.(1). 

Our findings, which are rather revealing, are summarized in Table I.  
Since the fluctuations were very large, we have only reported order of
magnitude results for the death rates. Pure diploids have the lowest death rate, 
an order of magnitude smaller than pure haploids. Mitotic (including the 
``episodic" case) sexuals are very close to diploids. For non-hereditary sexuals 
the rates are comparable to asexuals, but hereditary sexuals have a death rate
that is two orders of magnitude larger than asexuals. We infer from this
that the success of hereditary sexuals in lowering their mutation load is
due to the much greater rate at which they can select-out highly mutated
genes through death.

Thus it is seen that the clear advantage of sex over asex (haploid or
diploid) in our model can only become manifest with a sufficiently high 
frequency of
sexual reproduction, which is driven by the mutation rate
$\Gamma$, and moreover in situations which can discriminate between the
different adaptabilities (smaller typical $m$) of the different modes of
reproduction, such as time varying environments.~\cite{Volkan} This is
similar to the finding of Pekalski~\cite{Pekalski}, who has considered
environments (wildtypes) that vary over time, and found that the benefits
of sexual reproduction are enhanced by higher birthrates. For lower
birthrates, meiotic parthenogenesis, which can be compared to diploid
asex, has a slight advantage over sex, a difference from our results which
can be ascribed to the exponential fitness function he uses.

\section{Discussion}

This paper is a culmination of a series of studies where we have 
considered very stringent rules for the survival and propagation of diploid, 
sexual 
individuals in competition with haploids. 
 Here we have finally been able to show that a pair of simple, unicellular 
organisms 
who have accidentally converted to diploidy, and which subsequently engage in 
sexual reproduction, 
begetting one sexual offspring, can give rise to a population of sexual types 
which totally 
take over a finite population, provided they are also allowed to multiply by 
mitosis, on an 
equal footing with the haploids in the population.  They engage in sex when the 
going gets tough,
that is, when the number of their expressed deleterious mutations exceeds a 
certain number. This success seems to  vindicate the hypothesis of 
JSM~\cite{Jan1}, that sex could have 
been  a mechanism of last resort when simple organisms were faced with 
extinction.

It is interesting to pose the question of what happens if we do not allow the 
diploids 
to clone themselves, but, on the other hand, allow the gametes, under special 
conditions, 
to enter a haploid phase where they multiply by mitosis.  Eventually these 
haploids will 
be allowed once more to fuse and give rise to diploids.  Within the present 
scheme, 
since the haploids convert to diploidy only at the threshold, $m=4$, the gametes 
of the 
diploid individuals have $m\ge 4$, i.e., they are not viable.  Therefore the 
extension of the haploid phase calls for a modification 
of the rules in such a way as to allow the gametes  to survive nevertheless, 
e.g., by 
forming ``spores." 

One way the parameters of the present model could be modified is to take the 
threshold for survival to be different (greater than) the threshold for the 
conversion to (or practice of) sex.  Then, if a haploid gamete were to be 
considered as a member of the haploid population, and subjected to the same 
rules for selection and reproduction, it would seem as if one would get 
haploid phases of arbitrary duration.  However, upon closer inspection we see 
that this is not true:  Since the gamete comes from a diploid organism which 
engages in sex due to an excess mutational load, it already has deleterious 
mutations in excess of (or at least equal to)  the sex threshold. (For the 
diploid organism to have $m$ expressed mutations, both of its genetic 
strings must have at least that many). So, according to the rules which we have 
adopted, the abbreviated haploid cycle is automatically selected, since within 
this scheme the gamete would immediately seek a mate.  The lengthening of the 
haplophase would call for further evolution, e.g., such mutations as would lead 
to the enhancement of the fitness of the haploid gametes with $m$ in excess of 
the sexual threshold (say by the formation of spores, to survive 
in an environment which has become disfavorable [3,4]). Since this will provide 
a mechanism for the diploids once more to make 
up for the reduction in their numbers due to the single-offspring mating rule, 
we expect this 
HDC to win out over the haploid, asexual population.

To understand the seemingly contradictory of results from similar models 
investigating the phenomenon of the selection of sex as the dominant mode of 
reproduction, it is of interest to compare our model with that of 
Redfield~\cite{Redfield} and  Cui et 
al.,~\cite{Cui}.  The former employs a more complex system in which there is 
sexual differentiation between males and females, with the additional feature of 
a much greater mutation rate for males than for females, eventually making 
meiotic parthenogenesis safer than sex.  Redfield has used three different 
``selection functions" (survival probabilities), falling off exponentially (no 
``epistasis," or correlation between the effects of successive mutations) 
quadratically (some positive epistasis) or as a step function (extreme positive 
epistasis) with $m$\cite{Kondrashov}.  The fact that the step-function 
(``truncation") survival probability alone leads to a much better fitness resulting from sex, tells us that the form of the survival probability which is adopted 
determines the outcome very strongly.   The step function favors sex 
because it does not punish small deviations from the wildtype, thereby allowing 
genetic diversity, while it severely penalizes large 
deviations, helping eliminate highly mutated individuals.  

This insight enables us to understand the results of Cui et al. who have 
considered a diploid population reproducing either asexually or sexually, 
depending upon a fixed probability. There is a  constant mutation rate.  The 
survival probability is chosen to be of the ``independent mutation," i.e., 
exponential type. These authors find that the diploid population accumulates 
such a  large number of deleterious mutations (which however do not get 
expressed), that sexual reproduction, i.e., the random pairing of gametes from 
different parents,  results in a disasterous reduction in the fitness of the 
offspring. Moreover, sexually reproducing diploid populations are susceptible to 
invasion by asexually reproducing ones, while the converse is not true, i.e., 
asexually reproducing diploid populations are not invaded by sexually 
reproducing ones.  These results change drastically in favor of sex when they 
introduce cell senescence, which means a cell can clone itself only a certain 
number of times before it stops dividing and dies, unless it 
engages in sexual reproduction, which resets the senescence clock.

We would like to argue that in the model of Cui et al., introducing cell 
senescence is very much like turning on a  ``truncation like"  survival 
function, which, as in our model, also controls the switch to sexual 
reproduction.  In the presence of a constant mutation rate, the number of times 
a cell has cloned is another way to measure its (mean) mutation load, which, in 
an asexual haploid population of course gets directly expressed, whereas 
for diploid individuals hides behind the dominant unmutated alleles.  
Introducing a cell senescence threshold, beyond which the individual is either 
killed off or has to engage in sexual reproduction, is therefore equivalent, in 
an average way,  to the JSM criterion, applied to a diploid population.  Seen in 
this way, it is very gratifying that the Cui et al. model in fact corraborates 
our findings regarding the greater fitness of diploid populations 
that frequently engage in sex, given a step-function like survival probability.  

Very recently, two variants of the Redfield model, which assign ``harsher selection'' to males (rather than relatively higher mutation rates) have been considered~\cite{Siller,Agarwal}. These studies  both find that the mutational load on the whole population decreases as a result, and that the relative fitness of sexual females is increased by more than enough to compensate for the twofold cost of sex in anisogamous populations, at least for extremely high values of the average mutation rate.  Agarwal finds that this can happen even in the absence of synergystic epistasis, in his model. Wilke et. al~\cite{Wilke} have meanwhile found that high mutation rates select for low replication rates and flat regions of the fitness surface.

It should be noted that the Fermi-type fitness function (1) adopted
here\cite{Jan1,Thoms,Orcal,Tuzel} extrapolates, for finite temperatures
(smaller values of $\beta$) between highly synergistic (step function) and
independent (exponential) survival probabilities\cite{Lenski} as a function of 
$m$. 
A recent study by Peck and Waxman indicates~\cite{Peck} that competition for 
limited resources can
lead to synergy between successive mutations, leading to truncation, or
step function-like survival probabilities, which they also find favors
sexual populations.

In this paper we have presented numerical results for a model involving
unicellular organisms, which may shed light on how sex and diploidy
emerged and established a foothold in the protozoon world. Many
different mechanisms have so far been proposed via which sex may prove
advantageous or otherwise, in more highly evolved organisms. We believe
that care should be taken while proposing any single mechanism, such as an
adaptation to resist infestation by parasites~\cite{Lively,SaMartins}, for
the preferance of sex by organisms that range from the unicellular
eukaryotes to trees or human beings~\cite{Hamilton}.  More complex
organisms may have elaborated much more complex survival mechanisms and
behaviorial patterns, which stabilize or destabilize already evolved
traits.

{\bf Acknowledgements}

We thank Asl\i han Tolun and Canan Tamerler Behar for many useful discussions.
One of us (A.E.) gratefully acknowledges partial 
support from the Turkish Academy of Science.

\end{document}